\documentclass[conference]{IEEEtran}
\IEEEoverridecommandlockouts
\usepackage{cite}
\usepackage{amsmath,amssymb,amsfonts}
\usepackage{algorithmic}
\usepackage{graphicx}
\usepackage{textcomp}
\usepackage{xcolor}
\usepackage{hyperref}
\usepackage{subcaption}
\usepackage{booktabs}
\usepackage{tabularx}
\usepackage{multirow}

\def\BibTeX{{\rm B\kern-.05em{\sc i\kern-.025em b}\kern-.08em
    T\kern-.1667em\lower.7ex\hbox{E}\kern-.125emX}}

\usepackage{fancyhdr}
\fancypagestyle{firstpage}{
  \fancyhf{}
  \fancyhead[C]{\footnotesize \textcolor{black}{Accepted for publication at \href{https://percom.org/accepted-papers-demo/}{IEEE Percom Demo Track 2026}}}
}

\begin{document}
\title{Visual Insights into Agentic Optimization\\of Pervasive Stream Processing Services}

% \author{
% Anonymous Anonymous\and\inst{1,*} \and
% Anonymous Anonymous\and\inst{1} \and
% Anonymous Anonymous\and\inst{2} \and
% Anonymous Anonymous\inst{1,2}}
% %
% \authorrunning{Anonymous et al.}
% \institute{Anonymous
% \and
% Anonymous\\
% * Corresponding author contact: \email{anonymous} 
% }

\author{
\IEEEauthorblockN{
Boris Sedlak\IEEEauthorrefmark{1},
Víctor Casamayor Pujol\IEEEauthorrefmark{1},
Schahram Dustdar\IEEEauthorrefmark{1}\IEEEauthorrefmark{2}
}
\IEEEauthorblockA{\IEEEauthorrefmark{1}
Distributed Intelligence and Systems-Engineering Group,
Universitat Pompeu Fabra, Barcelona, Spain\\
Email: \{boris.sedlak, victor.casamayor, schahram.dustdar\}@upf.edu}
\IEEEauthorblockA{\IEEEauthorrefmark{2}
Distributed Systems Group,
TU Wien, Vienna, Austria}
% \IEEEauthorblockA{{\footnotesize \Letter} \hspace{1pt} Corresponding author: boris.sedlak@upf.edu}
}

\maketitle
\thispagestyle{firstpage}

\begin{abstract}
Processing sensory data close to the data source---often involving Edge devices---promises low latency for pervasive applications, like smart cities. This commonly involves a multitude of processing services, executed with limited resources; this setup faces three problems: first, the application demand and the resource availability fluctuate, so the service execution must scale dynamically to sustain processing requirements (e.g., latency); second, each service permits different actions to adjust its operation, so they require individual scaling policies; third, without a higher-level mediator, services would cannibalize any  resources of services co-located  on the same device. This demo first presents a platform for context-aware autoscaling of stream processing services that allows developers to monitor and adjust the service execution across multiple service-specific parameters. We then connect a scaling agent to these interfaces that gradually builds an understanding of the processing environment by exploring each service's action space; the agent then optimizes the service execution according to this knowledge. Participants can revisit the demo contents as video summary and introductory poster, or build a custom agent by extending the artifact repository.
\end{abstract}


\section{Introduction}
\label{sec:introduction}

Sensory data is used for fueling and optimizing pervasive applications, from autonomous driving~\cite{liu2021vehicular} to smart cities~\cite{de_donno_foundations_2019}.
This is supported by the growing computational power of embedded devices and Edge servers that support low-latency processing close to the data source.
The precise requirements \textit{how} this processing must be done are specified through Service Level Objectives (SLOs); real-time applications---like point cloud mapping~\cite{liu_point_2021}---might specify a maximum target latency.
Yet, resources on Edge servers are limited, whereas client demand is fluctuating; this inevitably leads to situations where resources do not suffice to satisfy SLOs across multiple competing clients and applications. 
To ensure SLO fulfillment, autoscaling solutions---like Kubernetes \cite{kubescaler2023vpa}---have specialized in adjusting applications according to varying demand; however, the default mechanism is provisioning additional resources.
% , which cannot be guaranteed outside of Cloud centers.
Also, it cannot be assumed that computation can be offloaded to nearby devices~\cite{quattrocchi_autoscaling_2024_short}. As the context changes dynamically and services cannot rely on predefined mechanisms (e.g., offloading or resource scaling), the processing services must autonomously find actions that optimize their SLO fulfillment.

To facilitate the transition to flexible and context-aware autoscaling of processing services, we developed a two-fold approach~\cite{sedlak_multi-dimensional_2025_url}: MUDAP---our \underline{Mu}lti-\underline{D}imensional \underline{A}utoscaling \underline{P}latform---supports fine-grained adjustments of the service execution and the allocated resources; notably, this permits dynamic adjustments to service-specific parameters, like the size of Machine Learning (ML) models or input tensors. Second, we presented RASK, a scaling agent that uses \underline{R}egression \underline{A}nalysis of \underline{S}tructural \underline{K}nowledge to interpret the effect of different parameter assignments on the SLO fulfillment, and then infer optimal scaling actions. Together, they enable flexible processing services that scale different parameters according to the context---a behavior called \textit{multi-dimensional elasticity}~\cite{sedlak_towards_2025_short}.
Thus, services can trade off less critical aspects (e.g., client experience) to sustain critical SLOs (e.g., latency).

This demo first introduces the architecture of MUDAP and RASK; next, we design a scaling agent that uses these interfaces to optimize the performance of three stream processing services, co-located on an Edge device. To provide insights into this operation, we visualize the agent's understanding of the processing environment and show how its internal model and the SLO fulfillment improve parallel. We complement this with an \textbf{introductory poster}~\cite{sedlak2026_poster} for quickly conveying the high-level idea; additionally, we host the \textbf{demo application} at a public URL, provide a \textbf{video summary}~\cite{sedlak2026_video} to it, and share an \textbf{artifact repository}~\cite{sedlak2026_artifact} for revisiting the demo contents.

% The remaining paper is organized as follows: Section~\ref{sec:architecture} presents the solution architecture and Section~\ref{sec:demo-description} the combination of demo contents; Section~\ref{section:takeaways} summarizes key takeaways for demo participants. Section~\ref{sec:conclusion} concludes the demo paper.

\section{Solution Architecture}\label{sec:architecture}

In the following, we present an architecture for context-aware autoscaling of stream processing services, involving two components: MUDAP and RASK. MUDAP exposes service-specific parameters for fine-grained adjustments of the processing environment, while RASK uses the interfaces for interpreting and optimizing the environment. 
Later, we visualize the RASK agent's internal models and show how increasingly accurate world models improve decision-making.

\subsection{Autoscaling Platform (MUDAP)}

\begin{figure}[t]
    \vspace{-6pt}
    \centering
    \includegraphics[width=1.02\linewidth]{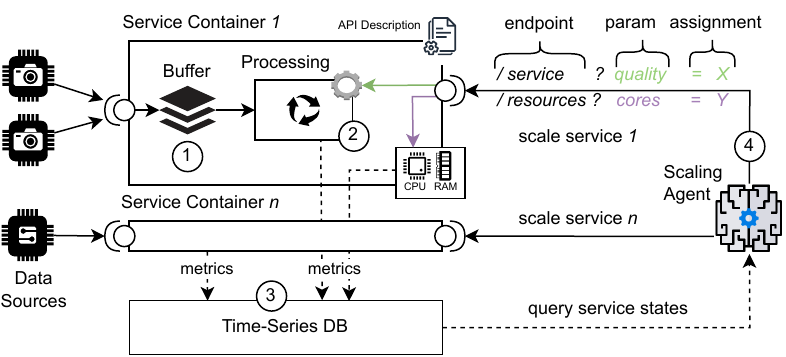}
    \caption{Architecture of the MUDAP platform~\cite{sedlak_multi-dimensional_2025_url}: sensor data is \textcircled{\small 1} buffered and \textcircled{\small 2} processed by containerized services; \textcircled{\small 3} service and container states (i.e., processing metrics) are collected in a time-series DB. Lastly, \textcircled{\small 4} a scaling agent interprets these states, develops a policy, and adjusts service configurations and their containers through a REST API. }
    \label{fig:mudap}
    \vspace{-6pt}
\end{figure}

% MUDAP
The MUDAP platform is introduced in Figure~\ref{fig:mudap} in four steps:
\textcircled{\small 1} It streams and buffers sensory data (e.g., video frames) at a nearby device, where multiple containerized processing services run.
\textcircled{\small 2} The data is processed, e.g., by running video inference.
\textcircled{\small 3} It continuously exports processing metrics to a time-series DB; this includes metrics about service executions (e.g., \textit{latency} or \textit{data quality}) and the associated resources (e.g., \textit{CPU limit}). These variables describe a service's state space; those variable that can be directly adjusted form the action space. For example, video resolution (i.e., \textit{data quality}) can be scaled dynamically. To invoke actions for a service (e.g., change its \textit{data quality}) we offer a REST API in the container.
\textcircled{\small 4} It optimizes service execution by coupling an agent to these interfaces. This allows arbitrary implementations of autoscalers; in our case, the RASK agent.

% which shows how containerized applications continuously process data and store metrics about the service and container states. A scaling agent, like the RASK agent, can consume these states and develop a scaling policy, which it executes by calling the respective endpoints in the MUDAP platform.

\subsection{Autoscaling Agent (RASK)}

\begin{figure}
    \centering
    \includegraphics[width=0.9\columnwidth]{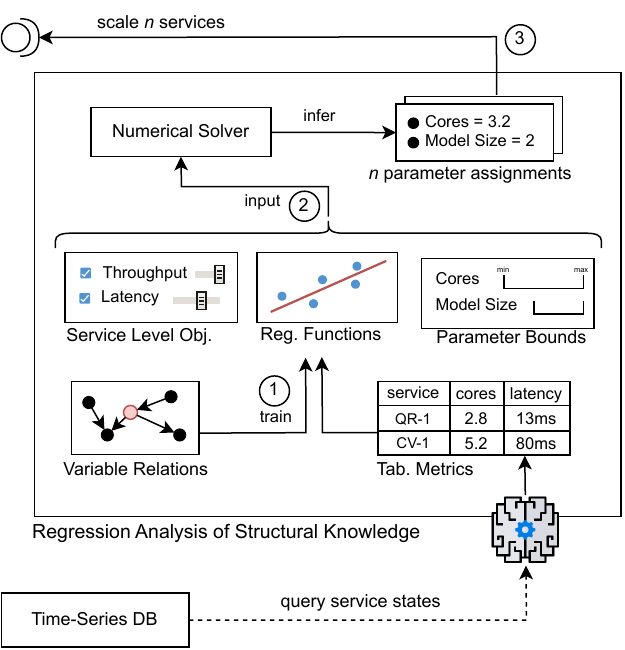}
    \caption{Conceptual sequence of RASK algorithm~\cite{sedlak_multi-dimensional_2025_url}: \textcircled{\small 1} create a tabular structure from time-series data and train regression functions; \textcircled{\small 2} supply functions, SLOs, and parameter bounds to numerical solver; \textcircled{\small 3} optimize parameter assignments for all monitored services and adjust values through MUDAP API.}
    \label{fig:rask-overview}
\end{figure}

% RASK
To optimize the execution of pervasive stream processing services, we present RASK alongside Figure~\ref{fig:rask-overview} in three steps:
\textcircled{\small 1} The agent models the behavior of the processing environment by fitting regression functions, using tabular metrics from the time-series data and domain knowledge about variable relations\footnotemark. This tells  the agent \textit{how} its interventions (e.g., changing \textit{data quality} or \textit{resources}) affect processing (e.g., \textit{inference latency}).
\textcircled{\small 2} For all services executed on the device, the agent collects the regression models, their SLOs, and the parameter bounds.
\textcircled{\small 3} The agent combines these factors into a global optimization function, uses a numerical solver to infer parameter assignments for all service, and adjusts the values through the API provided by MUDAP.

\footnotetext{For simplicity, we supply variable relations according to expert knowledge. However, the relation can equally be extracted through structural learning.}

\vspace{3pt}

By using this architecture, autonomous scaling agents can optimize the execution across multiple co-located processing services, thus supporting decentralized and robust operation.
In the demo setup we will bring up more details on \textit{how} the agent explores the solution space to develop an accurate understanding of the variable relations.
%
% A key aspect missing is \textit{how} the agent explores the solution space to develop an accurate understanding of the variable relations---we will explain this further in the demo setup.

% The scalability and applicability of RASK and MUDAP have been assessed in detail in \cite{sedlak_multi-dimensional_2025_url}; to summarize the findings, it showed that gradually adding more elasticity dimensions to the processing services improved the SLO fulfillment from \textbf{0.75\%} for 1 dimension, to \textbf{0.92\%} for 3 dimensions. However, as the number of services increases, the numerical solver of the RASK agent becomes a bottleneck: the SLO fulfillment reduces from \textbf{0.92\%} for 3 services to \textbf{0.85\%} for 9 services; similarly, the runtime increases, still 9 services could be managed with a mean runtime of 2s runtime, which easily fits the 10s autoscaling cycle.

% Use the platform created in \cite{sedlak_towards_2025_short,sedlak_multi-dimensional_2025}

\section{Demo Contents}
\label{sec:demo-description}

% In the following, we describe the demo's contents, the equipment required to hold the demo, and lastly, summarize the takeaways for demo participants.

% \subsection{Demo Contents}

The demo contains a multitude of contents that allow the participants to interact with it during the demo session, as well as afterwards. In the following, we present the contents in the same order in which participants should consume them.

\subsection{Introductory Poster}
\label{subsubsec:poster}

Ideally, participants read the \textbf{introductory poster}~\cite{sedlak2026_poster} before switching to the visual demo of the agentic service optimization.
Thus, they are already roughly aware of the problem domain, the solution's objectives and architecture, and the organization of the visual demo contents. Also, they can read at their own pace, ask questions that come up, or fall back to the poster when in doubt of the presented visual contents.

\subsection{Visual Demo of Operation}

Afterward, participants move to the visual demo of the scaling agent's operation, with three main parts:

\begin{figure}[t]
    \vspace{-10pt}
    \centering
\includegraphics[width=1.0\linewidth]{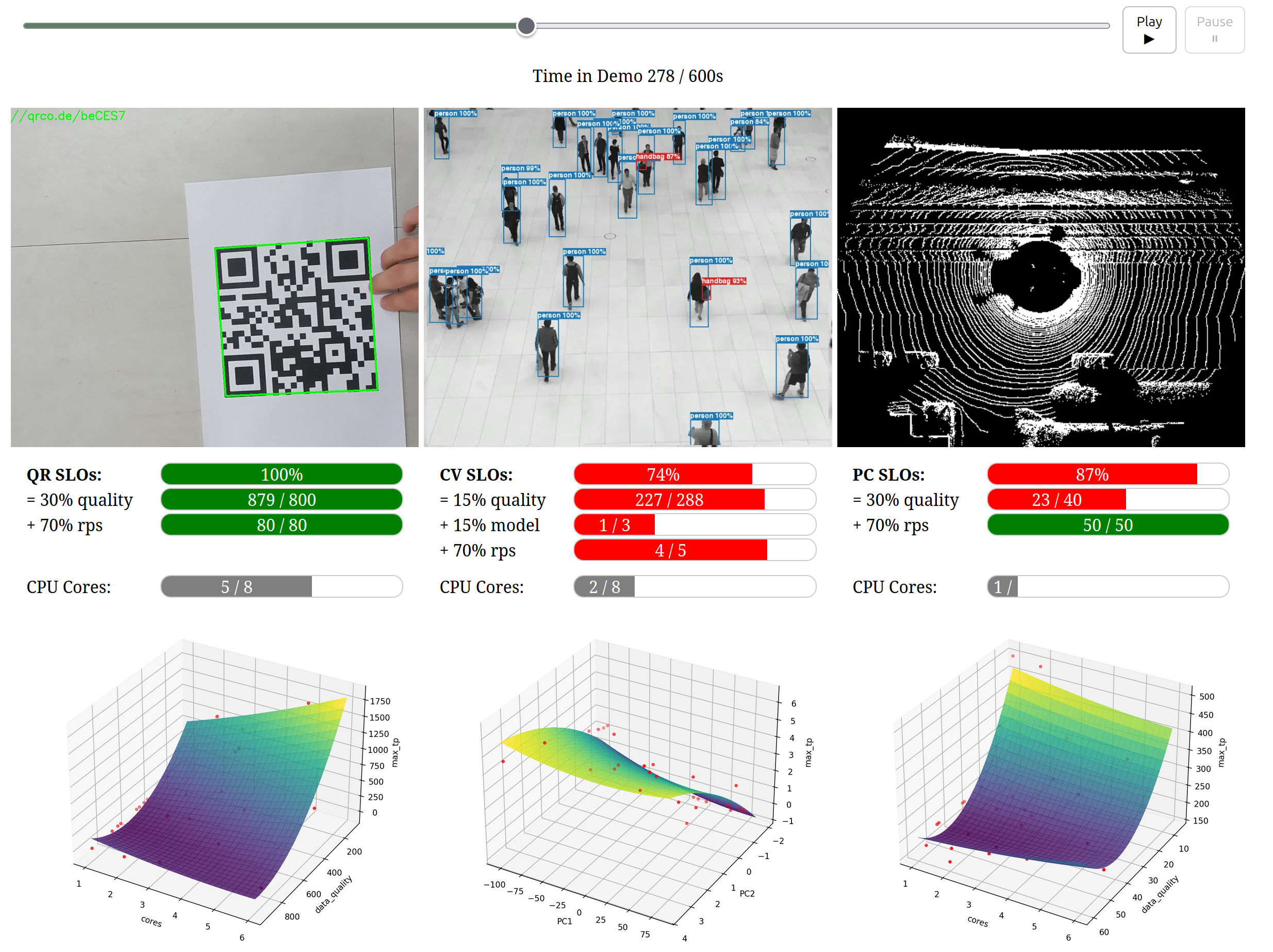}
    \caption{Snapshot of the visual demo: For three processing services, we display their current service output, their SLO fulfillment, and the regression model that the RASK agent learns through interventions with the autoscaling platform.}
    \label{fig:ui-snapshop}
    \vspace{-6pt}
\end{figure}

\subsubsection{Experimental Setup}

% \ins{describe the contents that should be known to understand the demo: (1) the services, (2) their SLOs, (3) their parameters, ()}

The visual demo covers an autoscaling agent that optimizes the execution of three stream processing services: we implement a QR code reader, a Computer Vision (CV) service using Yolov8, and a Point Cloud (PC) mapper; Figure~\ref{fig:ui-snapshop} shows an exemplary service output in the top row. To achieve SLO fulfillment, the scaling agent adjusts the resource allocation between services, and each service's parameters. As shown in Table~\ref{tab:slo-thresholds}, variables have viable bounds and a step size. For all three services, the agent can adjust the \textit{data quality} and the allocated CPU resources from a limited global budget of 8 \textit{cores}; the CV service also allows changing the Yolov8 \textit{model size}. Lastly, the \textit{completion} rate of processed items cannot be directly set, but depends on the other variables---this is the domain knowledge we supply to the agent. According to variables' importance to the system operation, we assign SLO thresholds and weights to them.

The demo covers a duration of 600s; in the first 300s, the scaling agent randomly explores the solution space to create an understanding of the processing environment, in the second 300s, the agent exploits the best configurations known. The agent operates in cycles---every 10s, it collects metrics, builds the regression models, and infers an autoscaling policy.

\begin{table}[t]
\setlength{\tabcolsep}{4.5pt}
\centering
\caption{Service variables for the QR, CV, and PC service; SLO targets and their \textbf{w}eights (i.e., importance) included. }
\begin{tabular}{cllcc|lc}
\toprule
\textbf{Service} & \textbf{Variable} & \textbf{Descript.} & \textbf{Bounds} & \textbf{Step} & \textbf{SLO} & \textit{w} \\
\midrule
\multirow{3}{*}{QR} 
    & \textit{cores}          & CPU quota     & $(0,8)$        & float  & --        & -- \\
    & \textit{data quality}   & Image size    & $[10^2,10^3]$  & $\pm 1$ & $\geq 800$ & 0.5 \\
    & \textit{completion}     & Rate finish   & $[0,1]$        & --     & $\geq 1.0$ & 1.0 \\
\midrule
\multirow{4}{*}{CV} 
    & \textit{cores}          & CPU quota     & $(0,8)$       & float  & --        & -- \\
    & \textit{data quality}   & Image size    & $[128,320]$   & $\pm 32$ & $\geq 288$ & 0.2 \\
    & \textit{model size}     & Yolov8[n/.]   & $[1,4]$       & $\pm 1$ & $\geq 3$   & 0.2 \\
    & \textit{completion}     & Rate finish   & $[0,1]$       & --     & $\geq 1$   & 1.0 \\
\midrule
\multirow{3}{*}{PC} 
    & \textit{cores}          & CPU quota     & $(0,8)$       & float  & --        & -- \\
    & \textit{data quality}   & Lidar range   & $[6,60]$      & $\pm 1$ & $\geq 40$ & 0.5 \\
    & \textit{completion}     & Rate finish   & $[0,1]$       & --     & $\geq 1$   & 1.0 \\
\bottomrule
\label{tab:slo-thresholds}
\end{tabular}
\vspace{-6pt}
\end{table}

\subsubsection{Visual Animation}

Figure~\ref{fig:ui-snapshop} shows the demo screen: on the top, the current time in the experiment and the control for the playback are displayed. To create a fluent animation, the playback is accelerated by $\times10$; hence, the entire demo lasts 60s, and every second the configuration and SLO fulfillment are updated. Each column shows the service output, the SLO fulfillment, and the agent's regression model---its understanding of how to achieve high \textit{completion} by adjusting the available parameters. Interested readers can access the public \textbf{demo application}~\cite{sedlak2026_demo_url} to track the agent's progress.

\subsubsection{Results}

The experimental demo shows how operating the RASK agent for 300s---equivalent to 30 interventions in the environment---suffices to improves global SLO fulfillment from 56\% to 98\%. This is considerably sample-efficient when compared with contemporary RL approaches, like Q-learning. In the later 300s, the agent keeps stable, high SLO fulfillment by adjusting the parameters along the Pareto front of optimal assignments; this behavior is also tracked in the video.

\subsection{Explanatory Video}

The \textbf{video summary}~\cite{sedlak2026_video} contains a technical explanation of the paper's methodology using the introductory poster, and a showcase of the visual demo progress. This allows conference participants and other researchers to revisit the contents.

\subsection{Artifact Repository}

The \textbf{artifact repository}~\cite{sedlak2026_artifact} (published under CC BY-NC-SA license) invites researchers to reuse the existing processing environment---including the three services---and couple their custom implementation of a scaling agent. Also, to support researchers in creating their own autoscaling platform, it shows how to expose arbitrary scaling parameters for their services.
%
 % \textbf{CC BY-NC-SA} license.

% \section{Demo Takeaways}
% \label{section:takeaways}

% % This demo combines different complementing contents that allow participants to interact with the autoscaling platform in their own pace. Participants can use the demo video to refresh their understanding of the demo contents, and the project repository to extend the autoscaling platform and scaling agent. 

% In summary, demo participants learn about (1) the necessity to develop alternative elasticity strategies for resource-restricted devices; the autoscaling platform gives them hands-on examples of possible strategies, and (2) the design and execution of custom scaling agents that can in few steps be connect to the platform.
% %
% Finally, by decomposing our own scaling agent, participants learn (3) how modeling the processing environment through regression functions can gradually optimize service execution. Through these contents, we hope to engage with other researchers to improve our methodologies and spark their interest.

\section{Conclusion}\label{sec:conclusion}

This paper presented a solution for context-aware autoscaling of stream processing applications under strict resource limits. Our approach lets scaling agents explore services' action space, build a model of the processing environment, and gradually optimize the service execution. Scaling agents run directly on the Edge device, thus increasing the autonomy and robustness of pervasive processing services.
We created a visual demonstration of the execution of our scaling agent: in under 30 iterations, it develops an accurate environmental model, fulfilling SLOs of three co-located services over 98\%. We accompany the demo with an introductory poster, an explanation video, and an artifact repository---researchers can revisit and extend these contents at their own pace.

% \section*{Required Equipment}

% We would kindly request a stand for the introductory poster and a large TV monitor for showing the demo screen.

\section*{Acknowledgment}
This work is partially supported by CNS2023-144359 and the European Union NextGenerationEU/PRTR under MICIU/AEI/10.13039/501100011033.
% \vspace{-5pt}

\bibliographystyle{IEEEtran}
\bibliography{boris,references}

@article{de_donno_foundations_2019,
	title = {Foundations and {Evolution} of {Modern} {Computing} {Paradigms}: {Cloud}, {IoT}, {Edge}, and {Fog}},
	volume = {7},
	issn = {2169-3536},
	shorttitle = {Foundations and {Evolution} of {Modern} {Computing} {Paradigms}},
	doi = {10.1109/ACCESS.2019.2947652},
	urldate = {2025-12-05},
	journal = {IEEE Access},
	author = {De Donno, Michele and Tange, Koen and Dragoni, Nicola},
	year = {2019},
	pages = {150936--150948},
}

@article{liu_point_2021,
	title = {Point {Cloud} {Video} {Streaming}: {Challenges} and {Solutions}},
	volume = {35},
	issn = {1558-156X},
	shorttitle = {Point {Cloud} {Video} {Streaming}},
	doi = {10.1109/MNET.101.2000364},
	number = {5},
	urldate = {2024-05-28},
	journal = {IEEE Network},
	author = {Liu, Zhi and Li, Qiyue and Chen, Xianfu and Wu, Celimuge and Ishihara, Susumu and Li, Jie and Ji, Yusheng},
	month = sep,
	year = {2021},
	pages = {202--209},
}

@article{liu2021vehicular,
  title={Vehicular edge computing and networking: A survey},
  author={Liu, Lei and Chen, Chen and Pei, Qingqi and Maharjan, Sabita and Zhang, Yan},
  journal={Mobile networks and applications},
  volume={26},
  number={3},
  pages={1145--1168},
  year={2021},
  publisher={Springer}
}

@online{kubescaler2023vpa,
    author = {Kubescaker Labs},
    url = {https://kubernetes.io/blog/2023/05/12/in-place-pod-resize-alpha/},
    month = {May},
    year = {2023},
    urldate = {Jul, 25, 2025}
}

@article{quattrocchi_autoscaling_2024_short,
	title = {Autoscaling {Solutions} for {Cloud} {Applications} {Under} {Dynamic} {Workloads}},
	issn = {1939-1374},
	doi = {10.1109/TSC.2024.3354062},
	urldate = {2025-05-28},
	journal = {IEEE Transactions on Services Computing},
	author = {Quattrocchi, Giovanni and Incerto, Emilio and Pinciroli, Riccardo and Trubiani, Catia and Baresi, Luciano},
	month = may,
	year = {2024},
}

@inproceedings{sedlak_towards_2025_short,
	title = {Towards {Multi}-dimensional {Elasticity} for {Pervasive} {Stream} {Processing} {Services}},
	copyright = {All rights reserved},
	urldate = {2024-04-25},
	booktitle = {2025 {IEEE} {PerCom} {Workshops}},
	author = {Sedlak, Boris and Morichetta, Andrea and Raith, Philipp and Pujol, Victor Casamayor and Dustdar, Schahram},
	year = {2025},
}

@misc{sedlak_multi-dimensional_2025_url,
	title = {Multi-{Dimensional} {Autoscaling} of {Stream} {Processing} {Services} on {Edge} {Devices}},
	copyright = {All rights reserved},
	doi = {10.48550/arXiv.2510.06882},
	urldate = {2025-10-12},
	publisher = {arXiv},
	author = {Sedlak, Boris and Raith, Philipp and Morichetta, Andrea and Pujol, Víctor Casamayor and Dustdar, Schahram},
	month = oct,
	year = {2025},
    url={https://arxiv.org/abs/2510.06882},
}

@misc{sedlak2026_video,
	author = {Sedlak, Boris},
  title        = {{Explanation} {Video} for {Demo} {Showcase}},
  url          = {https://borissedlak.github.io/percom-demo-2026/video/}
}

@misc{sedlak2026_poster,
	author = {Sedlak, Boris},
  title        = {{Percom} {Demo} {Introductory} {Poster}},
  url          = {https://borissedlak.github.io/uploads/poster-percom-demo.pdf}
}

@misc{sedlak2026_artifact,
	author = {Sedlak, Boris},
  title        = {{Elastic} {Workbench} {Artifact} {Repository}},
  url          = {https://github.com/borissedlak/elastic-workbench/tree/percom-demo}
}

@misc{sedlak2026_demo_url,
	author = {Sedlak, Boris},
  title        = {{Elastic} {Workbench} {Public} {Demo} {Application}},
  url          = {https://borissedlak.github.io/percom-demo-2026/}
}

% \appendix
% % \section{Introductory Poster}
% \includepdf[pages=-]{poster_anonymized.pdf}
% \label{sec:appendix}

\end{document}